\documentclass[aps,prb,reprint,groupedaddress]{revtex4-1}
\usepackage{amsmath}
\usepackage{txfonts}
\usepackage{type1cm}
\usepackage{bm}
\usepackage{graphicx}

\begin{document}

\title{Meissner effect cannot be explained classically}
\author{Daijiro Yoshioka}
\affiliation{Department of Basic Science, The University of Tokyo, 3-8-1 Komaba, Tokyo 153-8902, Japan}

\date{\today}

\begin{abstract}
The Meissner effect is an important characteristic of superconductivity and is critical to distinguishing superconductivity from simply the absence of electrical resistance (perfect conductivity). In a recent paper published in American Journal of Physics, Ess\'en and Fiolhais claimed that the Meissner effect is explained by classical physics. [Am. J. Phys. {\textbf{80}} 164, (2012).] We claim it cannot be understood by classical mechanics and point out that their derivation of the Meissner effect by classical physics is based on an inadequate treatment of the magnetic field energy. A correct treatment of the magnetic field energy clarifies the need for quantum mechanics to understand the Meissner effect.
We stress that Meissner effect is energetically favorable due to the energy of condensation of the Cooper pairs.
The condensation of electrons into Cooper pairs is best understood as a quantum mechanical phenomenon.

\end{abstract}
\maketitle

\section{Introduction}

When we apply a magnetic field to a normal metal, the magnetic field penetrates the metal.
If the metal becomes superconducting when the metal is cooled, the magnetic field in the bulk of the superconductor may be expelled.
The magnetic field is expelled, if the strength of the field is lower than temperature-dependent critical field of the superconductor.
This expulsion of the magnetic field is known as the Meissner effect.\cite{meissner}  This phenomenon is distinct from that of perfect conductivity. In a classical system with zero resistance Lenz's law guarantees that the magnetic flux density cannot change in time.  Thus if a system becomes a perfect conductor below some critical temperature, its magnetic flux density will depend on its history; i.e. whether the external field was applied before or after cooling.  For sufficiently weak fields, superconductivity is a well-defined thermodynamic phase in which the equilibrium state of the system is independent of history.  The expulsion of the field contradicts with the Lenz law of the classical electrodynamics, and cannot be explained without quantum mechanics.  The expulsion of the magnetic field costs energy which is supplied by quantum mechanical condensation energy of the electrons into Cooper pairs.

It is also noteworthy that superconductors are classified into two types.
For type I superconductor, there is only one temperature-dependent critical field, $H_c(T)$.
If the strength of the magnetic field exceeds this value, the superconductivity is destroyed, and the magnetic field penetrates the bulk.
A schematic phase diagram of the type I superconductor is shown in Fig.~1.
For type II superconductor, there are two critical fields, $H_{c1}(T)$ and $H_{c2}(T)$.
When the magnetic field satisfies $H_{c1}(T)<H<H_{c2}(T)$, quantized magnetic fluxes penetrate the superconductor, forming an Abrikosov lattice.\cite{abrikosov}
The magnetic flux is quantized in units of the elementary flux quantum $h/2e$, where $h$ is the Planck constant and $e$ is the elementary charge.
This quantization clearly indicates that this phenomenon originates from the quantum mechanical effect.

\begin{figure}[h]
\includegraphics[width=8cm]{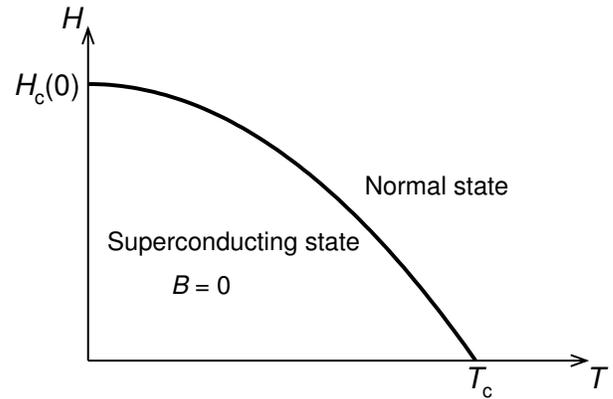}
\caption{A schematic phase diagram of a type I superconductor.
Strength of the  critical magnetic field $H_c(T)$ is shown by a thick line as a function of temperature $T$.
In the superconducting state below the line, applied magnetic field is expelled from the superconductor so that in the bulk of the superconductor the magnetic flux density $B$ is zero.
In the normal state, applied field penetrates the metal.
The transition temperature to the superconducting state in the absence of a magnetic field is shown as $T_c$.}
\label{fig:1.1}
\end{figure}

The magnetic field is expelled from the superconductor via a dissipationless supercurrent that flows at the surface of the superconductor.
This supercurrent flows such that the magnetic field created by this current just cancels the external magnetic field in the bulk of the superconductor.
The supercurrent is possible because of a complex order parameter which describes the macroscopic wave function of the condensed Cooper pairs.
The phase of the order parameter must be single valued in the bulk of the superconductor.
This condition makes dissipationless current in the superconductor possible in spite of the presence of impurities and phonons which give finite resistivity in the normal metal.
The realization of the supercurrent can only be explained quantum mechanically.
The relation between the complex order parameter and the supercurrent is supported by much experimental evidence including the Josephson effect.\cite{josephson}

The existence of the supercurrent, however, is not enough to explain the Meissner effect.
When a system crosses the phase boundary shown in Fig.~1 from normal side to superconducting side, the supercurrent begins to flow spontaneously, and the magnetic field is expelled.
In this process the magnetic flux in the superconductor changes from finite value to zero.
According to classical electromagnetism, Lenz's law works to resist the change in the magnetic flux.
To expel the magnetic field from the bulk of the superconductor against this law, it is required that the superconducting state in which the magnetic field is expelled has lower free energy than the state in which the magnetic field remains in the bulk.
In the absence of the magnetic field, the superconducting state has lower free energy than the normal state provided $T<T_c$, where $T_c$ is the transition temperature to the superconductivity.
In the presence of the magnetic field, superconducting state that expels the magnetic field competes with the normal state in which the magnetic field remains.
Superconductivity cannot be realized when the magnetic field penetrates the bulk, so the superconductor needs to expel the magnetic field.
However, expelling the magnetic field costs energy, so when the applied magnetic field becomes strong enough, the energy of the superconductor becomes higher than that of the normal state, and the superconductivity is destroyed.
This is the origin of the critical field.

The lowering of the energy in the superconducting state comes from the condensation energy of the electrons into Cooper pairs.
It is described by the BCS theory,\cite{bcs} which is entirely based on the quantum mechanics.
The temperature dependence of the critical field observed experimentally is quantitatively explained by the theory.
Thus, quantum mechanics is indispensable for the explanation of the Meissner effect.

In spite of all this, Ess\'en and Fiolhais\cite{essen} claimed that the Meissner effect can be explained by classical mechanics.\cite{essen2}
From our argument above, it is evident that their claim is incorrect.
It is however instructive to examine their claim and understand the flaw of their argument.
In the next section, we introduce their argument.  In Sec.III, we see how several strange conclusions are derived if we accept their argument.
In Sec.IV we clarify the flaw of their argument and find that they treat energy of the magnetic field incorrectly.
We need only classical electromagnetism to understand the flaw in their argument. 
No understanding of quantum mechanics or superconductivity is necessary to understand our arguments in this section.

\section{Argument by Ess\'en and Fiolhais}

Ess\'en and Fiolhais try to persuade us that Meissner effect is a property of a perfect conductor.
So let us forget the fact that the supercurrent can only be explained quantum mechanically, and try to examine whether or not a perfect conductor shows the Meissner effect.
Their argument is based on a description in chapter 1 of a famous textbook by de Gennes,\cite{degennes} where the London equation is derived using only classical mechanics. The derivation is not appropriate as we will see in Sec.IV, but we repeat the argument here.

For simplicity, the discussion will be limited to $T=0$.
It is assumed that total energy of superconductor in a magnetic field is given by
\begin{equation}
E=E_s+E_B+E_k,
\end{equation}
where $E_s$ is the condensation energy of electrons, which comes from quantum mechanical effects.
The second term,
\begin{equation}
E_B=\frac{1}{2\mu_0} \,\int\bm{B}^2 dV,
\end{equation}
is the energy of the magnetic field applied externally with $\mu_0$ being permeability of the vacuum,\cite{cgs}, $\bm{B} =\mu_0\bm{H}$ is the magnetic flux density with $\bm{H}$ being the applied magnetic field and
\begin{equation}
E_k = \int \frac{1}{2}n(\bm{r})m\bm{v}^2(\bm{r}) dV
\end{equation}
is the kinetic energy of the (super) current.
In this equation $m$ is the mass of an electron, $n(\bm{r})$ is the number density of electrons, and 
$\bm{v}(\bm{r})$ is average velocity of electrons.
This energy can be rewritten as follows.
The supercurrent density $\bm{j}(\bm{r})$ is expressed as 
\begin{equation}
\bm{j}(\bm{r}) =n(\bm{r})e\bm{v}(\bm{r})\,.
\end{equation}
It is also related to rotation of the magnetic flux density owing to Ampere's law that is valid for a steady current:
\begin{equation}
\nabla\times \bm{B}(\bm{r})  = \mu_0 \bm{j}(\bm{r}) .
\end{equation}
Using these equations the kinetic energy is expressed as
\begin{equation}
E_k=\frac{1}{2\mu_0}\int\lambda^2(\nabla\times \bm{B})^2\,dV,
\end{equation}
where $\lambda=\sqrt{m/\mu_0ne^2}$ is the London penetration depth.
Now de Gennes assumes $E_s$ to be a constant, and minimizes the energy $E_B+E_k$ with respect to the field distribution $\bm{B}(\bm{r})$, and obtains the London equation:
\begin{equation}
\bm{B}+\lambda^2\nabla\times(\nabla\times\bm{B})=0\,,
\end{equation}
as a condition to minimize the energy.
A solution gives a state in which supercurrent flows at the surface layer of the conductor, the thickness of the layer being of the order of $\lambda$,\cite{lambda} and the magnetic field in the bulk is eliminated.
This situation is identical to the state realized by the Meissner effect.
The essence of the argument is that the energy of the magnetic field $E_B$ is minimized by making $B=0$ in the bulk.
In this derivation quantum mechanics is not used at all.
Ess\'en and Fiolhais claim based on this argument that the Meissner effect can be explained classically.
However, this coincidence is superficial as we will see later.

\section{Strange consequences}

If this argument by de Gennes is correct we encounter various strange conclusions.
We point out some of them in this section.

(1) In the argument in Sec.II, the condensation energy of the superconductor has no role.
Thus there is no reason for the superconducting state to be destroyed by an arbitrarily strong magnetic field. 
It cannot explain the existence of the critical field $H_c$ above which the superconductivity is destroyed and the magnetic field penetrates the bulk.

Furthermore, unphysical latent heat must exist at the phase transition point.
Let us consider what happens at the phase transition point.
We apply a strong enough magnetic field to a sample, and cool it to a temperature near absolute zero.
The magnetic field is stronger than the critical field, so the magnetic field penetrates the sample, and the sample is in the normal state.
(See Fig.~1.)
Then we reduce the magnetic field.
At $H=H_c(T)$, the sample transitions to a superconductor, and the magnetic field is expelled.
Let us consider a cylindrical sample with radius $R$,  length $l$, and volume $V=\pi R^2l$.
The magnetic field is applied parallel to the axis of the cylinder.
We investigate what happens to the energy in Sec.II at the transition.
Before the transition, there is no current, and the magnetic field stays in the sample.
The energy of the magnetic field ${E_B}^{(n)}$ and the kinetic energy ${E_k}^{(n)}$ in the normal state are easily calculated using eq.(2) and eq.(3), respectively.
The answers are ${E_B}^{(n)}=(1/2\mu_0)B^2V$, and ${E_k}^{(n)}=0$.

Next we calculate ${E_B}^{(s)}$ and ${E_k}^{(s)}$ in the superconducting state.
The supercurrent flows at the side wall of the cylinder and cancels the magnetic field in the bulk.
The current flows in a thin skin layer of the order of $\lambda$.
Since $\lambda$ is typically small, of the order of $10^{-7}$\,m, we can consider that the magnetic field is almost completely expelled from the volume $V$ for a sample with $R \gg \lambda$.
Thus, ${E_B}^{(s)}=0$.
On the other hand, ${E_k}^{(s)}$ is not zero in this phase.
From the London equation eq.(6), we know that the magnetic field decays at the surface as $B\exp[(r-R)/\lambda]$ with $0 \le r \le R$ being the distance from the central axis of the cylinder.
Using this magnetic field, we obtain $|\nabla \times \bm{B}| \simeq (B/\lambda) \exp[(r-R)/\lambda]$.
Then $E_k^{(s)}$ is calculated by eq.(6) as
\begin{align}
E_k^{(s)}
&=
\frac{B^2}{2\mu_0}\int \exp[2(r-R)/\lambda] dV\nonumber\\
&\simeq
\frac{B^2}{2\mu_0}\int_0^R  2\pi Rl \exp[2(r-R)/\lambda]dr \nonumber\\
&=
\frac{B^2}{2\mu_0}\pi Rl\lambda
=\frac{B^2}{2\mu_0}\,\frac{\lambda}{R}\,V\,.
\end{align}
In the second line we used the fact that the integrand has non zero value only at $r \simeq R$.
Gathering these results we find that energy in the normal state $E^{(n)}=E_B^{(n)}+E_k^{(n)}=(1/2\mu_0)B^2V$ is much larger than that in superconducting state
$E^{(s)}=E_B^{(s)}+E_k^{(s)}=(1/2\mu_0)B^2V(\lambda/R)$,
the ratio being $E^{(n)}/E^{(s)}=R/\lambda \gg 1$.
The difference of these energies and the condensation energy $E_s$ that is neglected in Sec.II must be compensated at the transition point in the form of a latent heat of the phase transition. 
However, such a latent heat is not observed experimentally.
Actually, the latent heat of a phase transition comes from change in the entropy, so it must be zero at $T=0$ owing to the third law of thermodynamics.
The argument of the previous section has no experimental support.

(2) A perfect conductor has no mechanism to dissipate energy.
Thus there is no process to bring the system to the thermal equilibrium state in which the free energy takes its minimum value.
The state in which $E_{B}=0$ in the bulk is never realized, if we start from a state in which $E_B>0$.

(3) In the derivation of the London equation, the fact that the metal is in superconducting state is not used at all.
Therefore, if this derivation is correct, any metal must show the Meissner effect.\cite{dissipation}

(4) Let us consider a long cylindrical bulk superconductor.
When the external magnetic field is parallel to the axis of the cylinder, supercurrent flows along the surface of the cylinder, 
and the field in the bulk is canceled.
This situation is like a solenoid used to cancel the magnetic field inside.
Now because the current flows only within a thin region at the surface of the cylinder, nothing will change when a hole is introduced along the axis of the cylinder.
Namely, we will have the same situation when we replace the bulk cylinder by a hollow tube;
$E_k$ stays the same and $E_B$ is independent of whether the space is filled with metal or not.
However, experimentally the bulk cylinder and the hollow tube behave quite differently.
For the hollow tube the magnetic field is not expelled from the hole in the tube.
The field remains as quantized flux with value quite close to that expected from a normal metal tube.
The argument in Sec.~II cannot explain this experiment.

(5) As we stated in (4) the bulk superconductor can be replaced by a superconducting solenoid whose leads are connected together. 
The supercurrent flows in the solenoid and cancels the magnetic field inside.
We can easily calculate the value of the current flowing.
For a solenoid with winding number $n$ per unit length, $I_0=B/n\mu_0$.
Now what happens when the superconductor connecting the leads is replaced by a normal conductor.
This replacement can easily be done by raising the temperature locally.
We do the operation at time $t=0$.
In this case, the current $I$ decays in time according to the equation of the classical electrodynamics.
\begin{equation}
L \frac{d I}{d t}+RI=0,
\end{equation}
with the solution
\begin{equation}
I(t)=I_0\exp\left(-\frac{R}{L}t \right),
\label{eq:8}
\end{equation}
where $L$ is the self-inductance of the solenoid, and $R$ is the resistance of the normal part.
As the current decays, the magnetic field penetrates inside the solenoid.
At the same time Joule heating occurs in the normal part.
The total amount of the Joule heat generated is easily calculated.
It is same as the amount of the magnetic field energy that is introduced inside the solenoid.\cite{note1}
According to the argument in the previous section, the energy of the magnetic field is increased, since the solenoid is filled with the field.
At the same time the same amount of the energy is dissipated as Joule heat.
Where does this energy come from?
How can the state realized as the minimum energy state change spontaneously to a higher energy state?
We cannot answer these questions, if we accept the argument in the previous section.

\section{The origin of the flaw}

As we have seen in the previous section, the argument by Ess\'en and Fiolhais brings consequences incompatible with experiments.
Here we explain what is wrong in their argument.
We need only knowledge of the electromagnetism to pinpoint their flaw. 

The essence of their argument is that the Meissner effect occurs to lower the magnetic field energy $E_B$ in the volume of the conductor.
They believe that this energy can be reduced by flowing current at the surface of the conductor.
However, this is wrong.
As we stated in Sec.I, we need energy to cancel the magnetic field in the conductor.
We calculate the energy needed to remove the magnetic field in several ways, and show that it is positive.

We consider a situation in which a magnetic flux density $B$ is created by an external magnet.
The field is assumed to be almost uniform in the space in which we do experiments.
We bring a solenoid whose length is $l_A$, the cross section is $S$ and winding number per unit length is $n$.
We use this solenoid several times, so we name it as solenoid A.
We place this solenoid A parallel to the external field $B$, and try to cancel the magnetic flux density inside the solenoid, the volume of which is $V=l_AS$.
It is noted that when the current flowing in the solenoid is zero, inner space of the solenoid contains magnetic energy
\begin{equation}
E_B=\frac{1}{2\mu_0}B^2V\,.
\label{eq:9}
\end{equation}
Let us assume for simplicity that the solenoid A is made of a perfect conductor, so the resistance is zero.
The self inductance of the solenoid A is
\begin{equation}
L=\mu_0n^2V\,.
\end{equation}

In order to cancel the magnetic field inside, we need to flow current $I_0=B/\mu_0n$.
This current does not flow spontaneously.
We need to apply electromotive force ${\cal E}(t)$ to the solenoid.
The equation for the current is
\begin{equation}
{\cal E}(t) = L\frac{d}{dt}I(t)\,.
\end{equation}
The work done by the electromotive force is
\begin{align}
W &= \int_{t_1}^{t_2} {\cal E}(t)I(t) dt= \int_{t_1}^{t_2} LI(t)\frac{d}{dt}I(t)dt\nonumber\\
&=
\frac{1}{2}L\left(I(t_2)^2-I(t_1)^2\right)
=
\frac{1}{2\mu_0}B^2V\,.
\label{eq:11}
\end{align}
Here we used the condition that $I(t_1)=0$, and $I(t_2)=I_0=B/n\mu_0$.
The amount of energy required 
is the same as that of the magnetic field stored in the volume $V$, but energy must be supplied externally.
Namely, in order to remove magnetic field energy $E_B$ from volume $V$, we need to input the same amount of energy to the solenoid.
Where does sum of these energies, $B^2V/\mu_0=W+E_B$, go?
We will explain the destiny of this energy later.

We can calculate the energy another way.
A magnetic field exerts force on an electric current.
Thus, the current in a solenoid is pushed by the magnetic field.
This force is known as Maxwell stress, and the wall of the solenoid is pushed by pressure $B^2/2\mu_0$, the direction of which is from the high field side to low field side.
Now we consider an infinitesimally thin superconducting solenoid placed parallel to the external magnetic field $B$.
Since the self inductance of the solenoid is infinitesimally small, we can flow current to cancel the field inside without energy.
Suppose that the solenoid is made of plastic material, and we can expand the radius while keeping the field inside to be zero.
Since the wall of the solenoid is pushed from outside by the pressure $P=B^2/2\mu_0$, work is needed for this expansion.
The work to realize a solenoid with inner volume $V$ is $PV=(B^2/2\mu_0)V$, the same work as before.

We have shown that even though a finite amount of magnetic energy is stored in space with density $B^2/2\mu_0$, we need the same amount of energy to eliminate this energy.
For the Meissner effect, this energy is supplied by the condensation energy of the superconducting state.
We cannot extract the magnetic energy and utilize it without supplying extra energy.
Ess\'en and Fiolhais erroneously argued that the Meissner effect occurs to reduce this energy $E_B$ without work $W$.
As we have seen in Sec.~III, their argument brings conclusions incompatible with experiments.

Finally, we explain the energy balance.
To eliminate external magnetic field $B$ from the volume $V$, we need work $W=(B^2/2\mu_0)V$.
This work and the energy of the magnetic field $E_B$ must be absorbed somewhere.
We can show that the energy is absorbed in the electromagnet that is creating the external magnetic field $B$, or as potential energy of the magnet when the field $B$ is due to a permanent magnet.
For simplicity, let us consider the case in which the field $B$ is created by another solenoid (which we label solenoid B) in which current $I_B$ is flowing.

The field at solenoid A should be proportional to $I_B$, so we write the relation as $B=\alpha I_B$.
Since the magnetic flux that penetrates solenoid A is $\Phi=nl_ASB=\alpha nl_ASI_B$,
the mutual inductance between solenoid B and solenoid A is calculated to be $M=\Phi/I_B=\alpha nl_AS=\alpha nV$.
In the course of our increasing the current $I_A$ in solenoid A to cancel the magnetic field $B$, electromotive force ${\cal E}_B$ is induced in solenoid B:\cite{recipro}
\begin{equation}
{\cal E}_B=M \frac{dI_A}{dt}=\alpha nV\frac{dI_A}{dt}\,.
\end{equation}
The work done by this electromotive force on the constant current $I_B$ is
\begin{equation}
W_B=\int_{t_1}^{t_2} {\cal E}_B I_B dt=\alpha nV I_BI_0\,,
\end{equation}
where $I_A(t_1)=0$ and $I_A(t_2)=I_0$ is used.
Now we use the relation $\alpha I_B=B$ and $nI_0=B/\mu_0$.
Then we obtain $W_B=(B^2/\mu_0)V$, which is just sum of the work we used to cancel the magnetic field in the solenoid $W$, Eq.~(\ref{eq:11}), and the magnetic energy $E_B$, Eq.~(\ref{eq:9}), which existed in the volume.
Our argument is consistent with classical electrodynamics.

The situation in which the magnetic field inside the solenoid A is canceled by the current $I_0$ is a higher energy state, as we have emphasized several times.
When we make the solenoid A to have finite resistance $R$, the work $W$ spent to realize the situation is returned from the source of the magnetic field, i.e. from the power source of the solenoid B, as Joule heat generated in solenoid A.
Actually, it is calculated as follows,
\begin{align}
Q&=\int_{0}^{\infty} RI(t)^2dt=\int_{0}^{\infty} RI_0^2\exp\left(-2\frac{R}{L}t \right)\nonumber\\
&=\frac{L}{2}I_0^2=\frac{1}{2\mu_0}B^2V\,,
\end{align}
where $I(t)=I_0[-(R/L)t]$, Eq.~(\ref{eq:8}), $L=\mu_0n^2V$, and $I_0=B/n\mu_0$ are used.

Since the state where the magnetic field is expelled has higher magnetic field energy, the Meissner effect is possible only when the condensation energy to the superconducting state is larger than the magnetic field energy.
This condition determines the critical field $H_c$.
For ordinary metals the magnetic susceptibility is quite small, so the free energy of the normal metal, $F_n$ is almost independent of the magnetic field applied.
If the free energy of the superconducting state in the absence of the magnetic field is $F_s$, the critical field $H_c=B_c/\mu_0$ is determined by the condition,\cite{tinkham, kittel}
\begin{equation}
F_s + \frac{1}{2}\mu_0 H_c^2 V=F_n\,.
\end{equation}
In this equation $V$ is the volume of the superconductor, and $F_n$ is the free energy when the superconductivity is destroyed by the magnetic field.
This relation has been experimentally confirmed.

\section{Conclusions}

As is commonly understdoood, the Meissner effect requires quantum mechanics for its explanation. 
We have reviewed established evidence for this in the introduction.
This fact was challenged by Ess\'en and Fiolhais in a recent publication.
We have explained their argument in Sec.II, and have seen in Sec.III that if we accept their argument, unphysical phenomena are anticipated.
We clarified what is wrong in their argument.

\begin{acknowledgments}
D.Y. acknowledges the hospitality of the Aspen Center for Physics, which is supported by the National Science Foundation Grant No. PHY-1066293.
He thanks Steven Girvin and Allan MacDonald for critical reading of the manuscript.
\end{acknowledgments}


\begin{thebibliography}{99}

\bibitem{meissner} W. Meissner and R. Ochsenfeld, ``Ein neuer Effect bei eintritt der Superleitf\"ahigkeit,'' Naturwisse. {\textbf{21}}, 787-788 (1933).

\bibitem{abrikosov} A. A. Abrikosov, Zh. Eksperim. i Teor. Fiz. {\textbf{32}}, 1442-1452 (1957) [English transl., ``On the Magnetic Properties of Superconductors of the Second Group,'' Soviet Phys. JETP {\textbf{5}}, 1174-1182 (1957)].

\bibitem{josephson} B. D. Josephson, ``Possible new effects in superconductive tunneling,'' Phys. Letters {\textbf{1}}, 251-253 (1962).

\bibitem{bcs} J. Bardeen, L. N. Cooper, and J. R. Schrieffer, ``Theory of Superconductivity,'' Phys. Rev. 408, 1175-1204 (1957).

\bibitem{essen} H. Ess\'en and M.C.N. Fiolhais, ``Meissner effect, diamagnetism, and classical physics -- a review'', Am. J. Phys. {\textbf{80}} 164-169, (2012).


\bibitem{essen2} In addition to try to derive the Meissner effect classically, they try to disprove the Bohr-van Leeuwen theory.
This is because once we admit the theorem we conclude that the Meissner effect cannot have a classical explanation.
They point out several situations where classical systems show magnetic response as evidence for the break down of the theorem.
However, their argument is based on incorrect understanding of the magnetic field energy.
Furthermore, none of their evidence is magnetic response in the thermal equilibrium state.
They are confusing dynamical response of the system with thermal response of the system.
Their arguments do not disprove the theorem which states that classical calculation gives a partition function that is independent of the external magnetic field, so that the susceptibility becomes zero in the thermal equilibrium state.





\bibitem{degennes} P.G. de Gennes, \textit{Superconductivity of Metals and Alloys} (Adison-Wesley, Reading, MA, 1989).


\bibitem{cgs} We use SI instead of cgs-system used in Refs.~\onlinecite{essen} and \onlinecite{degennes}.




\bibitem{lambda} For typical superconductors, London penetration depth is smaller than $10^{-7}$\,m.

\bibitem{dissipation}
It might be argued that a normal metal has dissipation process and inclusion of the process invalidates the argument in Sec.II.
However, it is easily seen that such an effect does not change the situation.
If the resistance of the metal stops the current and favors the state with magnetic field, inclusion of the dissipative process must lower the energy of the system compensating the cost of the energy $E_B$.
Such an energy must be originated in the region where the current flows.
However, since the magnetic energy $E_B$ is proportional to the system volume, but the volume current flows is proportional to the surface area, inclusion of finite resistance cannot change the energy minimum condition in Sec.II.


\bibitem{note1} The calculation of the energy is given in Sec.IV.

\bibitem{recipro} The reciprocity theorem of the electromagnetic induction is used here.

\bibitem{tinkham} M. Tinkham, \textit{Introduction to Superconductivity}, 2nd ed. (Dover, Mineola, N.Y. 1996), pp.22-23.

\bibitem{kittel} C Kittel, \textit{Introduction to solid state physics}, 8th ed. (J.Wiley, Hoboken, N.J., 2005), pp.270-273.

\end{thebibliography}
\end{document}